\documentclass[12pt]{article}

\usepackage{graphicx}

\begin{document}
\begin{center}
{\Large\bf \boldmath On Finite Width of Quark Gluon Plasma Bags} 

\vspace*{6mm}
{ K. A. Bugaev }\\      
{\small \it  Bogolyubov Institute for Theoretical Physics,
Kiev, Ukraine \\      
         }
\end{center}

\vspace*{6mm}

\begin{abstract}
Within an exactly solvable model I discuss an  influence  of the  medium dependent finite width of QGP bags on their equation of state.  It is shown that  inclusion of such a width   allows one to naturally resolve two conceptual problems of the QGP statistical description.
On the basis of  the proposed simple kinetic model for a sequential  decay  of heavy QGP bags 
formed in high energy elementary particle collisions
it is argued that by measuring the energy dependence of  life time of these bags it is possible 
to distinguish  the case of  critical point  existence  from the case of  tricritical point.
\end{abstract}

\vspace*{6mm}

{\it 1. Introduction. --}
A lot of experimental and theoretical efforts is aimed  to determine the equation of state (EoS)
of the strongly interacting matter.  Despite the great achievements of these efforts \cite{EOSachiv}  even the bulk properties of  the quark-gluon plasma  (QGP)  EoS  are not well known. 
Thus, such important characteristics as the 
mean volume and life time of QGP bags formed in heavy ion collisions  have not caught a necessary attention yet. 
It is clear, however, that  right these quantities may put some
new bounds on the spacial and temporal properties of the QGP created in 
high energy collisions. 
As shown in  \cite{Wong:HBT}  it is possible to naturally resolve the HBT puzzles at RHIC energies, if one assumes that  the QGP consists of droplets of finite (mean) size. 
On the other hand  the short life time of heavy QGP bags found recently 
within the Hagedorn-Mott resonance model \cite{Blaschke:04}
and within 
the finite width model (FWM)  \cite{FWM:08, Reggeons:08} may not only play an  important role in  all thermodynamic and hydrodynamic phenomena of the strongly coupled QGP  matter, 
but may also explain 
the absence of strangelets  \cite{Strangelets:06}  or, more generally,  why the finite QGP bags cannot be observed
at energy densities typical for hadronic phase \cite{FWM:08} (see below). 
Therefore, an investigation of the  mass and volume distributions along with  the life time of the QGP bags and the corresponding   consequences for both the experimental observables and 
theoretical  studies   is  vitally   necessary for heavy ion phenomenology. 
The present paper   is devoted to a   discussion   of  these problems in the framework of 
the FWM.

{\it 2. The Finite Width Model. --}
The FWM employs the most convenient way to study the phase structure  of  any statistical  model  by  analyzing  its  isobaric partition \cite{Bugaev:04a, Bugaev:05c, Bugaev:07,CGreiner:06} and to find the  rightmost singularities of this partition. Hence,  I  assume that after the Laplace transform  the  FWM grand canonical  partition  $Z(V,T)$ generates the following 
isobaric partition:
\begin{eqnarray}\label{Zs}
\hspace*{-0.25cm}\hat{Z}(\lambda,T) \equiv \int\limits_0^{\infty}dV\exp(-\lambda \, V)~Z(V,T) =\frac{1}{ [ \lambda - F(\lambda, T) ] } \,, 
\end{eqnarray}
where the function $F(\lambda, T)$ is a  generalized partition
\begin{eqnarray}\label{FsI}
F(\lambda, T) = \int\limits_{0}^{\infty}dv\hspace*{-0.1cm}    \int\limits_{0}^{\infty}ds \hspace*{-0.1cm}    \int\limits_{0}^{\infty} 
 \hspace*{-0.1cm}dm~\rho(m,v,s)\exp(-\lambda \, v) \phi(T,m) ~
\end{eqnarray}
of bags of mass $m$, volume $v$ and surface $s$ defined by 
their  mass-volume-surface spectrum $\rho(m,v,s)$. The partition (\ref{FsI}) is a generalization
of the statistical ensembles with fluctuating extensive quantities discussed recently in  
\cite{GorenHauer:08}. Note that one could also introduce in  (\ref{FsI}) the perimeter fluctuations,
which may play an important role 
for small hadronic bubbles  \cite{Madsen} or for cosmological phase transition  \cite{Ignat:1},
but we neglect it because   the curvature term has not been seen in such well established models like the Fisher droplet model (FDM) \cite{Fisher:67,Elliott:06}, the  statistical multifragmentation model (SMM) \cite{Bondorf:95,Bugaev:00} and many other 
cluster systems discussed in \cite{Bugaev:05c, Bugaev:07,CGreiner:06,Elliott:06}. A special analysis of the free energy of 2- and 3-dimensional  Ising clusters, using the Complement 
method \cite{Complement}, did not find any traces of the curvature term (see a detailed discussion in \cite{Bugaev:07}). 
Such a result   is  directly related  to  the QGP bags because 
quantum chromodynamics (QCD)  is expected to be in the same universality class \cite{misha} as 
the 3-dimensional Ising model whose clusters were analyzed  in \cite{Complement}.

The  thermal  density
of  bags of mass $m$   and  a unit  degeneracy   
is given by  
\vspace*{-0.15cm}
\begin{eqnarray}\label{Bolzmann}
\phi (T, m)    \equiv  \frac{1}{2\pi^2} \int\limits_0^{\infty}\hspace*{-0.0cm}p^2dp~
\exp{\textstyle \left[- \frac{(p^2~+~m^2)^{1/2}}{T} \right] } 
 =  \frac{m^2T}{2\pi^2}~{ K}_2 {\textstyle \left( \frac{m }{T} \right) }\, .
\end{eqnarray}

\vspace*{-0.25cm}
\noindent
It is convenient to divide 
the mass-volume-surface spectrum 
into  the  discrete mass-volume spectrum of light hadrons and the continuum contribution
of heavy resonances $ \rho(m,v)$
\vspace*{-0.15cm}
\begin{equation}\label{FsHQ}
 \rho(m,v,s) =  \sum_{j=1}^{J_m}\, g_j \,\delta(m-m_j)\delta(v-v_j)\,\delta(s) +  \Theta(v -V_0) \Theta(m -M_0)  \delta(s - a_s v^\kappa) \rho(m,v) \rho_1 (s) \,,
\end{equation}

\vspace*{-0.25cm}
\noindent
The first term on the right hand side (r.h.s.) of (\ref{FsHQ}) represents the contribution of a finite number of low-lying
hadron states up to mass $M_0 \approx 2 $ GeV \cite{FWM:08}. This function has no $\lambda$-singularities at any temperature $T$ and can generate only a simple pole of the isobaric partition, whereas  the mass-volume spectrum of the bags  $\rho(m,v)$ on the r.h.s of  (\ref{FsHQ})  is chosen to 
generate an essential  singularity $\lambda_Q (T) \equiv p_Q(T)/T$ which defines  the QGP  pressure $p_Q(T)$.  For simplicity here I consider the matter with zero baryonic charge. 

The  continuous part of the spectrum $\rho(m,v,s)$ introduced in  \cite{FWM:08} is parameterized as
\begin{eqnarray}\label{Rfwm}
& \rho (m,v) =   \frac{ N_{\Gamma}}{\Gamma (v) ~m^{a+\frac{3}{2} } }
 \exp{ \textstyle \left[ \frac{m}{T_H}   -   \frac{(m- B v)^2}{2 \Gamma^2 (v)}  \right]  } \,, \quad
 {\rm and} \quad 
  \rho_1 \left(\frac{v^\kappa}{a_s}\right) = \frac{f (T)}{ v^{b}}~ \exp{\textstyle \left[  -  \frac{\sigma(T)}{T} \, v^{\kappa}\right] }\,.
\label{R1fwm}
 \end{eqnarray}

\vspace*{-0.25cm}
\noindent
As it is seen from (\ref{Rfwm}) the mass spectrum $\rho(m,v)$ has a Hagedorn like parameterization and  the Gaussian attenuation  around the bag mass
$B v$ ($B$ is the mass density of a bag of a  vanishing width) with the volume dependent  Gaussian  width $\Gamma (v)$ or width hereafter. 
I will distinguish it from the true width defined as 
$\Gamma_R = \alpha \, \Gamma (v)$ ($\alpha \equiv 2 \sqrt{2 \ln 2}\,$).
It is necessary to stress  that the Breit-Wigner attenuation  of  a resonance mass cannot be used in the spectrum 
(\ref{Rfwm}) because in case of finite width it would lead to a divergency of the mass integral in (\ref{FsI}) above  $T_H$  \cite{FWM:08,Reggeons:08}.

The normalization factor 
obeys the condition
$N_{\Gamma}^{-1}~ = ~ \int\limits_{M_0}^{\infty}
\hspace*{-0.1cm} \frac{dm}{\Gamma(v)}
\exp{\textstyle \left[  -   \frac{(m- B v)^2}{2 \Gamma^2 (v)}  \right] }
$.
The constants  $a > 0$ and  $b > 0$ define the Fisher exponent  $\tau \equiv  a + b$ \cite{FWM:08, Reggeons:08} (also see later). 

{\it 3. Important Features of the FWM Spectrum. --}
The  spectrum  in  (\ref{R1fwm}) contains the surface free energy (${\kappa} = 2/3$) with the $T$-dependent 
surface tension which is parameterized as 
$\sigma(T) = \sigma_0 \cdot
\left[ \frac{ T_{c}   - T }{T_{c}} \right]^{2l + 1} $  ($l =0, 1, 2,...$) \cite{Bugaev:07, Bugaev:04b},
where  $ \sigma_0 > 0 $ can be a smooth function of temperature.  For $T$ not above  the  tricritical temperature $T_{c}$ such a parameterization  is justified by the usual  cluster models 
like the FDM \cite{Fisher:67,Elliott:06} and SMM \cite{Bondorf:95,Bugaev:00}, whereas 
the general case for any  $T$   can be derived from the surface partitions of the Hills and Dales model \cite{Bugaev:04b}.  Note that  the Hills and Dales model \cite{Bugaev:04b}
explicitly  accounts for all possible surface deformations which correspond to the same cluster  volume and, therefore, it is another example of the statistical partition with fluctuating 
extensive quantity, which in this case  is cluster surface.

In Ref.  \cite{Bugaev:07} it was rigorously proven that at low baryonic densities 
the first order deconfinement phase transition degenerates into a cross-over just because of 
negative surface tension coefficient for $ T > T_{c} $. The other  consequences of the present 
surface tension  parameterization and the discussion of the absence of the curvature free energy in  (\ref{R1fwm}) can be found in Refs. \cite{Bugaev:07, Complement, PoS:06}. 

The power  $\kappa < 1$ which describes the bag's effective  surface is a constant that,  in principle, can differ from the typical FDM and SMM value 
$\kappa = \frac{2}{3}$ for which the coefficient $a_s$ is $a_s \equiv (36\, \pi)^\frac{1}{3}$.
This is so because  near  the deconfinement phase transition  region  the QGP  has  low density and, hence, 
like in the low density  nuclear matter \cite{Ravenhall},  
the non-spherical bags (spaghetti-like or lasagna-like \cite{Ravenhall})  can be  favorable
(see also \cite{Madsen,Ignat:1} for  the bubbles of complicated shapes).
A similar idea of  ``polymerization" of gluonic quasiparticles was introduced recently 
\cite{Shuryak:05a}. 

Note that in contrast to the continuous part of the spectrum (\ref{FsHQ}) its discrete part does not contain the surface free energy because  according to the present days status of the statistical 
model of hadron gas this  is not necessary \cite{HG}.

The spectrum (\ref{Rfwm}) has a simple form, but is rather general since both the width $\Gamma (v)$ and the bag's mass density $B$ can be medium dependent. In \cite{FWM:08} it is shown that the  FWM has no contradiction, if  $\Gamma(v) \equiv \Gamma_1 = \gamma v^\frac{1}{2}$ only ($\gamma = const$ of $v$).

For large bag volumes ($v \gg M_0/B > 0$) the normalization  factor $N_\Gamma$   can be
found to be $N_\Gamma \approx 1/\sqrt{2 \pi} $.   Similarly, one can show that  for heavy free bags  ($m \gg B V_0$, $V_0 \approx 1$ fm$^3$ \cite{FWM:08},
ignoring the  hard core repulsion and thermostat)
\vspace*{-0.15cm}
\begin{eqnarray}\label{Rm}
& \rho(m)  ~ \equiv   \int\limits_{V_0}^{\infty}\hspace*{-0.1cm} dv\,\rho(m,v) ~\approx ~
\frac{  \rho_1 (\frac{m}{B}) }{B ~m^{a+\frac{3}{2} } }
\exp{ \textstyle \left[ \frac{m}{T_H}     \right]  } \,.
\end{eqnarray}

\vspace*{-0.2cm}
\noindent
It originates in   the fact that  for heavy bags the 
Gaussian  in (\ref{Rfwm}) acts like a Dirac $\delta$-function for
 $\Gamma_1$. 
Thus, the Hagedorn form of  (\ref{Rm}) has a clear physical meaning and, hence, it  gives an additional argument in favor of the FWM. 

Similarly to (\ref{Rm}), it is possible to  estimate the width of heavy free bags  averaged over bag volumes and get  $ \overline{\Gamma(v) } \approx  \Gamma_1 (m/B) = \gamma \sqrt{m/B}$.
Thus, 
 the mass spectrum of heavy free QGP bags 
must be the Hagedorn-like one with the property that  heavy resonances should  have  the large  mean width  because of which  they  would be hard to be observed. 

The FWM allows one to express  the pressure of large bags in terms of their most probable mass and width. Comparing  the high and low temperature FWM pressures  \cite{FWM:08, Reggeons:08} with the lattice QCD data \cite{LQCD:1, LQCD:2, LQCD:3},  it was possible to estimate  the  minimal  resonance width  at zero temperature
$\Gamma_R (V_0, T=0) \approx  600 $ MeV and the width  at the Hagedorn temperature 
$\Gamma_R (V_0, T=T_H)   = \sqrt{12}\,\Gamma_R (V_0, T=0) \approx 2000$ MeV. 
It was also found that these values of the width are almost independent of the number of the lattice 
QCD elementary degrees of freedom  \cite{Reggeons:08}. 
Clearly, so  large  widths can naturally     explain the huge deficit of the heavy hadronic resonances in the 
Particle Data Group compared to the exponential mass spectrum used to describe the QGP EoS.  
Applying the same  line of arguments to the strangelets,
I conclude  that, if their mean volume is a few cubic fermis or larger, 
they  should survive for a  very short time. Such a conclusion
 is similar to the results of  Ref. \cite{Strangelets:06}  
predicting  an instability of  the strangelets.

Also it is  remarkable that at the temperatures  below the half of the Hagedorn one the  QGP bag pressure 
of the FWM 
acquires the linear $T$ dependence, i.e. $p^-(T < 0.5 T_H)   =  {\textstyle  - T\frac{ B^2}{2\, \gamma^2 } }$, which is clearly seen  in the recent lattice QCD data \cite{LQCD:3}  in the range $T \in [202.5; 419.09]$ MeV \cite{Reggeons:08}. 

As shown in \cite{Reggeons:08}   the relation between the resonance width and the mean mass of
the FWM  bags at high temperatures obeys the upper bound for the 
Regge trajectory asymptotic behavior found in \cite{Trushevsky:77}, whereas a similar relation 
at low temperatures exactly corresponds to lower bound for the Regge trajectory asymptotic form \cite{Trushevsky:77}. 

{\it 4. The Life Time of a Protofireball. --} 
The found FWM width values  allow one to get the rough estimates of the life time of the 
protofireball suggested in \cite{Protofireball}  to explain the hadron multiplicities measured in the elementary  particle collisions at high energies. The  microcanoncal analysis of the thermostatic
properties of  heavy resonances with  exponential mass spectrum \cite{Moretto:06, PoS:06} teaches 
us two principal  facts: first, even a single heavy resonance with the exponential mass spectrum imparts 
the Hagedorn temperature to any other hadron being in a thermal contact with it, 
and, second, the splitting of a single heavy resonance into several heavy pieces with mass above $M_0$ practically does not  alter the latter conclusion.
These two facts allow us to greatly simplify a treatment of the sequential decay 
of the heavy  QGP bags  formed in the elementary  particle collisions at high energies.
Indeed, the first fact allows one to consider the decay products with the mean 
kinetic energy which corresponds to the Hagedorn temperature (i.e. $3\, T_H$ for pions
and $3/2\, T_H$  for heavier particles), and the second fact enables us to study
the decay of several QGP bags independently of each other.  
Moreover, these both facts combined with the low particle densities formed in the 
elementary  particle collisions  allow one to neglect  the treatment of daughter hadrons  which may absorb on the decaying bags.

To simplify the problem I study the two particle decays only  and consider the evolution  of the heaviest QGP bag. The assumption of two particle decay is not too restrictive because it is possible to effectively account for three, four and more particle decays by representing them as the two particle sequential decays with shorter life-time. 
Therefore, in the rest frame of decaying bag of mass $M$
the mean  change of mass of the heaviest of two daughter particles of energies $E_1$ and $E_2$  is $\Delta M \approx M - \overline{\max (E_1, E_2)_M} = 
\overline{\min (E_1, E_2)_M}$, where the bar means the averaging over all possible combinations which obey the energy conservation.  Then the mass evolution equation for the heaviest QGP bag can be cast as
\vspace*{-0.2cm}
\begin{equation}\label{dMdt}
\frac{\Delta M}{\Delta \,\, t}~ \approx ~ - q\,  \Gamma_M~ \overline{\min (E_1, E_2)_M} \,,
\end{equation}

\vspace*{-0.25cm}
\noindent
where $\Delta \,\, t$ is the  time change,  $\Gamma_M =  \Gamma_R (V_0, T=0) 
\sqrt{\frac{M}{M_0} } \equiv  \Gamma_0 \sqrt{\frac{M}{M_0} } \approx 600 \cdot \sqrt{\frac{M}{M_0} }$ MeV is an average resonance width in the vacuum. Here the constant $q = \frac{\ln N}{\ln 2}$  accounts for the mean number of daughter particles  $N >2$ in a decay. 

The mass distribution of the heaviest bag can be constructed  from  the auxiliary function
\vspace*{-0.25cm}
\begin{equation}\label{OmegaM}
\Omega_M (E_1) \equiv  N_\Omega ~\rho( E_1) \int\limits_{E_{min}}^M d E_2 ~\rho(E_2) ~\delta\left(1 - \frac{E_1 + E_2}{M} \right) \, ,
\end{equation}

\vspace*{-0.25cm}
\noindent
where the density of states of hadrons of energy $E$ is denoted as $\rho(E)$, the Dirac delta function accounts for the energy conservation, and the normalization factor 
$N_\Omega$ is given by 
\vspace*{-0.2cm}
\begin{equation}\label{NOmega}
1 =  \int\limits_{E_{min}}^M d E_1 ~ \Theta(E_1 - M/2)~ \Omega_M (E_1)~ +
\int\limits_{E_{min}}^M d E_2 ~ \Theta(E_2 - M/2)~ \Omega_M (E_2)
  \, .
\end{equation}
Here the first (second) term corresponds to the fact that the fragment of energy $E_1$ ($E_2$) is the heaviest one, and  $E_{min} \approx 3 \, T_H$ is the minimal energy of the lightest  fragment.  Using (\ref{NOmega}), one finds  an averaged minimal mass of the daughter fragment as 
\vspace*{-0.2cm}
\begin{equation}\label{minM}
\overline{\min (E_1, E_2)_M} \approx  \hspace*{-0.3cm}
 \int\limits_{M/2}^{M - E_{min}} \hspace*{-0.3cm} d E_1 \, \rho(E_1)\, [M - E_1]\,
 \rho (M - E_1) 
 \left[
 \int\limits_{M/2}^{M - E_{min}} \hspace*{-0.3cm}d E_1 \, \rho(E_1)\,  \rho (M - E_1) \right]^{-1}
 \hspace*{-0.25cm}.
\end{equation}
 
\vspace*{-0.1cm}
\noindent
In principle the density of states should be defined from the convolution of the spectrum 
(\ref{FsHQ}) and Boltzmann density (\ref{Bolzmann}) with $T=T_H$. However,  to get a simple  analytic expression I employ just the Hagedorn mass spectrum and consider  all particles nonrelativistically. This can be done because  the experimental mass spectrum  $\rho_e (m)$ for $m \le M_0$ is well approximated by 
 $\rho_e (m) \approx C \left[\frac{M_0}{m} \right]^{\tau + 3/2} \exp ( \frac{m}{T_H}) $ \cite{Bron:04}. 
The energy conservation $E = m + \frac{3}{2}\, T_H + \tilde{\sigma}_0 \, m^\kappa $ 
relates the mean  energy  of daughter hadron and its mass. Neglecting the surface energy, I get the resulting energy spectrum of the daughter hadron  as $\rho(E) \approx C \left[\frac{M_0}{E} \right]^\tau \left[  \frac{e\, T_H\, M_0 }{2\, \pi} \right]^\frac{3}{2} $.
As one can see, the leading  exponentials cancelled each other and, thus,  $\overline{\min (E_1, E_2)_M} \approx  E_{min} \left[  \frac{M}{E_{min} } \right]^{2-\tau} 
 \left[  \ln \left( \frac{M}{E_{min} }  \right) \right]^{2\tau - 3} $ for $\tau \ge  1$. 
Hence the life time of the protofireball of mass $M_F$ created in the  high energy collision of elementary particles is
\vspace*{-0.05cm}
\begin{equation}\label{tM}
\hspace*{-0.15cm}
t_{\tau \neq \frac{3}{2}} (M_F)  \approx \frac{\Gamma_0^{-1}}{q\, (\tau - \frac{3}{2})} \sqrt{ \frac{M_0}{E_{min}} }
 \left[  \frac{E_{min}^{3/2-\tau} }{M_F^{3/2-\tau} }  -   \frac{E_{min}^{3/2-\tau} }{M_0^{3/2-\tau} }  \right], \,\,\, {\rm and} ~\,
t_{\tau = \frac{3}{2}} (M_F)  \approx \frac{\Gamma_0^{-1}}{q\,} \ln \left[ \frac{M_F}{M_0} \right].
\end{equation}
 
\vspace*{-0.1cm}
\noindent
Since for $\frac{3}{2} < \tau \le 2 $  at nonzero baryonic chemical potentials the deconfinement transition is of the first order and for  $\frac{4}{3} < \tau \le \frac{3}{2} $ it degenerates to the second order and in either  case there exists the tricritical point  \cite{Bugaev:07}, whereas there are some
arguments  that for $\tau > 2$ there should exist the critical point with the first order deconfinement \cite{Bugaev:05c}. Therefore,  I conclude that the measurements of the energy dependence of the  life time (hadronization time)  of the protofireballs created in the elementary
particle collisions may allow us to distinguish  the critical point  existence  from the tricritical point.

{\it 5. Conclusions. --}
Here I discuss some new ideas on the basic properties of the QGP EoS.
The main attention is paid to the role of the medium dependent width of heavy QGP bags. 
Their large width explains a huge deficit of experimental mass spectrum of  heavy hadronic resonances and enlightens some important thermodynamic aspects of the color confinement 
in finite systems. The proposed simple kinetic model for a sequential decay  of heavy QGP bags
formed in the elementary particle collisions at high energies allows one to distinguish  the case of  critical point  existence  from the tricritical one by measuring the energy dependence of the life time of these bags. Of course, the freeze-out process \cite{Bugaev:96} may some what change these conclusions,  but  the simple  kinetic arguments for  the nucleus-nucleus collisions 
\cite{ChemEq, Bugaev:02HC} teach  us that such changes are just a few per cent  only 
and, hence, they cannot affect the result obtained.




\end{document}